\documentclass[twocolumn]{revtex4-2}

\usepackage{slashed}
\usepackage{graphicx}
\usepackage{amsmath}
\usepackage{amssymb}
\usepackage{color}
\usepackage{float}
\usepackage[rflt]{floatflt}
\usepackage{mathtools}
\usepackage{dsfont}
\usepackage{subfig}

\newcommand{\Mvec}{{\rm \bf M}}
\newcommand{\Li}{{\rm Li}}
\newcommand*\pFqskip{8mu}
\catcode`,\active
\newcommand*\pFq{\begingroup
        \catcode`\,\active
        \def ,{\mskip\pFqskip\relax}%
        \dopFq
}
\catcode`\,12
\def\dopFq#1#2#3#4#5{%
        {}_{#1}F_{#2}\biggl[\genfrac..{0pt}{}{#3}{#4};#5\biggr]%
        \endgroup
}

\allowdisplaybreaks[4]

\begin{document}
\sloppy
\title[\boldmath $\mu$- and $q$-extension]
{\boldmath $\mu$- and $q$-extension of special functions in quantum field theory}%
\author{J. Bl\"umlein}
\affiliation{Deutsches Elektronen-Synchrotron DESY, Platanenallee 6, 15738 Zeuthen, Germany}
\affiliation{Institut f\"ur Theoretische Physik III, IV, TU Dortmund, 
Otto-Hahn Stra\ss{}e 4, 44227 Dortmund, Germany}
\author{A.M.~Gavrilik}
\affiliation{Bogolyubov Institute for Theoretical Physics, 14-B Metrolohichna str., Kyiv, 
03143, Ukraine}
\author{U.Y.~Lunha}
\affiliation{Bogolyubov Institute for Theoretical Physics, 14-B Metrolohichna str., Kyiv, 
03143, Ukraine}
\author{O.~Mykhailiv} 
\affiliation{Bogolyubov Institute for Theoretical Physics, 14-B Metrolohichna str., Kyiv, 
03143, Ukraine}
\author{C.~Schneider}
\affiliation{
Johannes Kepler University Linz, Research Institute for Symbolic
Computation (RISC),\\ Altenberger Stra\ss{}e 69, A-4040, Linz, Austria}

\vspace*{2mm}



\setcounter{page}{1}%

\begin{abstract}
\noindent
Analytic calculations up to single-scale quantities in perturbative quantum field theory 
result into special numbers and functions, which are generalizations as iterative integrals 
over special alphabets. In this way one obtains polylogarithms, Nielsen integrals, the 
iterated integrals over linear denominator terms, cyclotomic letters, letters induced by 
quadratic forms, and square-root valued letters. The inverse Mellin transform relates 
special nested sums to these quantities, which are harmonic sums, generalized harmonic 
sums, nested sums implied by quadratic forms, cyclotomic harmonic sums, and nested sums 
containing central binomials. We construct the $\mu$- and $q$-extensions of all these 
special functions. They are expected to occur in loop calculations within $\mu$- and 
$q$-extended quantum field theories via modified (anti)commutation relations. The 
$\mu$-extended special functions are quite different form the corresponding $q$-extended 
functions. While in the former case one observes class-preservation characterized by the 
polynomial $\mu$-parameter, except for the case of root-valued letters. In the latter 
case new higher transcendental functions are obtained.
\end{abstract}


\maketitle

\section{Introduction}
\label{sec:1}

\noindent
The $q$-extensions of functions, or the construction of $q$-basic functions, dates back 
to L. Euler to 1740 \cite{EULER2} in the theory of partitions and have been applied to
many mathematical functions later on. A further notable step were Heine's
$q$-hypergeometric functions \cite{HEINE2}. Later lots of orthogonal polynomials and other 
special functions, also higher hypergeometric functions, were $q$-extended, cf. e.g. 
Ref.~\cite{Blumlein:2026shz}.
In Quantum Mechanics and Quantum Field Theory the modified commutation relations
\cite{Arefeva:1995rpr}
\begin{eqnarray}
\label{eq:com}
a_\lambda(k) a_{\lambda'}^\dagger(k') - {q} a^\dagger_{\lambda'}(k') 
a_\lambda(k)
=  g_{\lambda \lambda'} \delta^{(3)}(k-k'),
\end{eqnarray}
imply $q$-extended structures. They are also related to Quantum Groups 
\cite{MANIN,KASSEL,CHARI,KS,Toller:2003yz}
and non-commutative space times 
\cite{CONNES,CONNES1,Kowalski-Glikman:2002eyl,Chaichian:1992fx}.
   
A second extension is the {$\mu$-extension}, 
Refs.~\cite{JANU,Gavrilik:2010xu},
\begin{eqnarray}
a a^\dagger - a^\dagger a &=& \varphi_\mu(N+1) - \varphi_\mu(N), 
\nonumber\\
\varphi_\mu(N)
&=& \frac{N} {1+\mu N}, \mu \in [0,1].
\end{eqnarray}
There are many physics applications both for the $\mu$- and $q$-extension. For 
recent summaries see Refs.~\cite{Blumlein:2026clg,Blumlein:2026shz} and work on the 
$\mu$-extension in Refs.~\cite{A1,A2,A3,A4,A5,A6,A7}.

In perturbative calculations in quantum field theory to higher orders different function
classes spanning the Feynman integrals in the zero- and one-variable cases have been 
found, cf.~\cite{Blumlein:2018cms}. 
In momentum fraction space, these are the 
polylogarithms \cite{Devoto:1983tc}, Nielsen integrals \cite{NIELSEN1,Kolbig:1983qt} , 
harmonic polylogarithms \cite{Remiddi:1999ew}, generalized harmonic polylogarithms
\cite{Moch:2001zr,Ablinger:2013cf}, cyclotomic harmonic polylogarithms 
\cite{Ablinger:2011te}, iterative  integrals induced by 
quadratic forms \cite{Ablinger:2021fnc}, and square-root valued iterative 
integrals \cite{Ablinger:2014bra}. The 
Mellin 
transform
\begin{eqnarray}
\Mvec[f(x)](N) = \int_0^1 dx x^{N-1} f(x), N \in \mathbb{N} \backslash \{0\}
\end{eqnarray}
relates these functions to the associated classes of nested sums
\cite{Vermaseren:1998uu,Blumlein:1998if,Ablinger:2013cf,Ablinger:2011te,
Ablinger:2021fnc,Ablinger:2014bra}.

We calculate the $\mu$- \cite{Blumlein:2026clg} and $q$-extensions 
\cite{Blumlein:2026shz} for these quantities. The different function spaces are defined 
in Section~\ref{sec:2}. In Section~\ref{sec:3} we present their $\mu$-extension and
in Section~\ref{sec:4} their $q$-extension. Section~\ref{sec:5} contains the 
conclusions.

\section{Higher function spaces in perturbative QFT}
\label{sec:2}

\noindent
We first consider non-$q-~(\mu-)$ extended perturbative quantum field theories for single 
scale processes, which obey {first order factorizing} differential or 
difference 
equations. \\
The results can be expressed by the following iterated integrals
\begin{eqnarray}
\label{eq:IT1}
G_{a_1,...,a_m}(x) &=& \int_0^x dy_1 f_{a_1}(y_1) \int_0^{y_1} dy_2 f_{a_2}(y_2)
.... 
\nonumber\\ 
& & ...\int_0^{y_{m-1}}
dy_m
f_{a_m}(y_m). 
\end{eqnarray}
The nested finite sums are defined by
\begin{eqnarray}  
S_{a,b_1,...,b_m}(N) = \sum_{k=1}^N s_a(k) S_{b_1,...,b_m}(k).
\end{eqnarray}
The Mellin transform relates $G_{a_1,...,a_m}(x)$ and $S_{a,b_1,...,b_m}(N)$.
The quantities $f_{a_m}(y_m)$ and $s_a(k)$ form special alphabets $\mathfrak{A}$ and
$\mathfrak{S}$.

In the following we list examples for the different-type nested sums and iterated 
integrals:

\begin{eqnarray}
\begin{array}{ll}
\vspace*{2mm}
\text{Harmonic polylogarithms:} & 
{\displaystyle \int_0^x \frac{dy}{y} \int_0^y \frac{dz}{1+z}} \\
\vspace*{2mm}
\text{Gen. harm. polylog.:} & 
\vspace*{2mm}
{\displaystyle  \int_0^x \frac{dy}{y} \int_0^y\frac{dz}{z-3}} \\
\vspace*{2mm}
\text{Cycl. harm. polylog.:} & 
{\displaystyle  \int_0^x \frac{dy}{1+y^2} \int_0^y\frac{dz}{1 -z + z^2}}\\
\text{root-valued iter. int.:} &
{\displaystyle  \int_0^x \frac{dy}{y} \int_0^y
\vspace*{2mm}
\frac{dz}{z\sqrt{1+z}}}
\end{array}
\nonumber
\end{eqnarray}

\begin{tabular}{ll}
\text{Harmonic sums:} & 
{$ \displaystyle \sum_{k=1}^N \frac{1}{k} \sum_{l=1}^k \frac{(-1)^l}{l^3}$} \\
\text{Gen. harmonic sums:} & 
{$ \displaystyle \sum_{k=1}^N \frac{(1/2)^k}{k} \sum_{l=1}^k \frac{(-1)^l}{l^3}$}\\
\text{Cycl. harmonic sums:} & 
{$ \displaystyle \sum_{k=1}^N \frac{1}{(2k+1)}\sum_{l=1}^k \frac{(-1)^l}{l^3}$}\\
\text{Binomial sums:} &
{$ \displaystyle \sum_{k=1}^N \frac{1}{k^2}   \binom{2k}{k} (-1)^k$} 
\end{tabular}

The iterated integrals for $x=1$ define the associated class of special numbers. We 
mention that the respective classes of iterated integrals and nested sums obey algebraic
shuffle resp. quasi-shuffle or stuffle relations, see  Ref.~\cite{Blumlein:2003gb}.

To construct the $\mu$- and $q$-extensions we consider the formal power series
expansion of the iterated integrals at $x=0$ in analytic form
\begin{eqnarray}
F(x) = \sum_{n=0}^\infty f[n] x^n.
\end{eqnarray}
The quantities $f[n]$ are given in terms of the associated nested sums.

The $\mu$- and $q$-extension are then obtained by the following replacements
\begin{eqnarray}
f[n] \rightarrow f[\{n\}_\mu],~~~~~~~f[n] \rightarrow f[\{n\}_q].
\end{eqnarray}
The modified values of $n$ are given by
\begin{eqnarray}
\label{eq:MU1}
\{n\}_\mu = \frac{n}{1 + \mu n},
~~~~\{n\}_q = \frac{1-q^n}{1-q}.
\end{eqnarray}
The $\mu$- and $q$-differential operators read
\begin{eqnarray}
\label{eq:MU2}
{\cal D}_x^{(\mu)} 
&=& \frac{d}{dx}\left[1 + \mu x \frac{d}{dx}\right]^{-1},
\\ 
{\cal D}_x^{q} f(x)  &=& \frac{f(x) - f(xq)}{(1-q)x}.  
\end{eqnarray}
The operator ${\cal D}_x^{(\mu)}$ is non-local and is understood in terms 
of a formal power series in the parameter $\mu$.

For some classes of iterated integrals one finds hierarchy relations, such as
for the multiple polylogarithms and the Nielsen integrals, which also hold for the
$\mu$- and $q$-extensions. For the polylogarithms one has
\begin{equation}
x {\cal D}_x \Li_m(x;\mu,(q)) = 
\Li_{m-1}(x;\mu,(q)),~~\text{etc.} 
\end{equation} 
\section{\boldmath The $\mu$-extension}
\label{sec:3}

\noindent
The $\mu$-extension of nested sums are implied by Eqs.~(\ref{eq:MU1}, \ref{eq:MU2}). For 
the single 
harmonic sums 
\begin{eqnarray}
S_{\pm m}(N;\mu) &=& \sum_{n=1}^N \frac{(\pm 1)^m}{n^m} 
\end{eqnarray} 
one obtains the $\mu$-extensions
\begin{eqnarray}
\label{eq:Sm}
S_{m}(N;\mu) &=& \sum_{n=1}^N \frac{1}{(\{n\}_\mu)^m} 
= \sum_{k=0}^m \binom{m}{k} \mu^k 
S_{m-k}(N), \nonumber\\ 
\end{eqnarray} 
for $m > 0,~~m \in \mathbb{N}$ and
\begin{eqnarray}
S_{-|m|}(N;\mu) &=& \sum_{n=1}^N \frac{(-1)^n}{(\{n\}_\mu)^{|m|}} 
\nonumber\\
&=& \sum_{k=0}^{|m|} 
\binom{|m|}{k} \mu^k
S_{-|m|+k}(N), 
\end{eqnarray} 
for $m < 0,~m \in \mathbb{Z}.$
The $\mu$-extension of the nested sums have a polynomial dependence in $\mu$,
reproducing lower weight nested sums of the same type. 
The term with the largest power of $\mu$ is $\propto N$. 

The most simple iterative integral is 
\begin{eqnarray}
\Li_k(x) = \sum_{n=1}^\infty \frac{x^n}{n^k}, x \in [-1,1]. 
\end{eqnarray}
Its $\mu$-extension is given by
\begin{eqnarray}
\Li_k(x;\mu) = \sum_{n=1}^\infty \frac{x^n}{(\{n\}_\mu)^k} = \sum_{l=0}^k \binom{k}{l}
\mu^l
\Li_{k-l}(x),
\end{eqnarray}

\noindent
for $x < 1$ with a similar structure as in the case of nested sums. 
In the large $x$-region one has
\begin{eqnarray} 
\label{eq:LImdiv} \Li_k(x;\mu) \propto \mu^k \frac{x}{1-x} - k \mu^{k-1}
\ln(1-x). 
\end{eqnarray} 
Note the divergent behavior for $x \rightarrow 1$.

Let us now consider the case of the $\mu$-extension of finite binomial sums.
The $\mu$-extended central binomial is given by
\begin{eqnarray}
\label{eq:BINsum}
\left. \binom{2n}{n}\right|_\mu = \frac{\displaystyle \prod_{k=n+1}^{2n}\left(k +
\frac{1}{\mu}\right)}{\displaystyle \prod_{k=1}^{n}\left(k + \frac{1}{\mu}\right)}
= \frac{\big(
        n + 1 
        +\frac{1}{\mu }
\big)_n}{\big(
        1+\frac{1}{\mu }\big)_n}.
\end{eqnarray}
An example for a $\mu$-extended finite binomial sum is 
\begin{eqnarray}
\lefteqn{\sum_{n=1}^N \frac{1}{4^n}
\left. \binom{2n}{n}\right|_\mu
=} \\ &&
\frac{ (1+2 \mu )}{4 (1+\mu )}
\pFq{3}{2}{1~,
\tfrac{3}{2} + \tfrac{1}{2 \mu}~,
2 + \tfrac{1}{2 \mu}~}{
2 + \tfrac{1}{\mu}~,
2 + \tfrac{1}{\mu}}{1}
\nonumber\\ &&
-\frac{ 4^{-N} \Gamma \big(
        1+\frac{1}{\mu }\big) \Gamma \big(
        1
        +\frac{1}{\mu }
        +2 N
\big)}{\Gamma \big(
        1
        +\frac{1}{\mu }
        +N
\big)^2}
\nonumber\\ &&
\times \left[-1 +\pFq{3}{2}{
1,
\tfrac{1}{2} + \tfrac{1}{2 \mu} +N,
1 + \tfrac{1}{2 \mu} +N}{
1 + N + \tfrac{1}{\mu},
1 + N+ \tfrac{1}{\mu}}{1} \right].
\nonumber 
\end{eqnarray}
The $\mu$-extension of the finite binomial sums maps to higher transcendental functions 
and is thus
{non-polynomial} in $\mu$.

\section{\boldmath The $q$-extension}
\label{sec:4}

\noindent
The $q$-extensions of nested sums and iterated integrals are non-polynomial. Examples are
\begin{eqnarray}
\hspace*{-3mm}
S_{3,-2}(n;q) &=& \sum_{k=1}^n \left[\frac{1-q}{1-q^k}\right]^3
\sum_{l=1}^k (-1)^l \left[\frac{1-q}{1-q^l}\right]^2,
\\
\hspace*{-3mm}
\Li_k(x;q) &=& \sum_{n=1}^\infty \frac{x^n}{\{n\}_q^k}
= \sum_{n=1}^\infty x^n \left[\frac{(1-q)}{(1-q^n)}\right]^k.
\end{eqnarray}
The $q$-extensions of the nested sums obey {$q$-difference} 
equations and those of the iterated integrals 
{$q$-differential} equations and {$q$-shift} relations.

An example for a $q$-difference equation is 
\begin{eqnarray}
&& \left(q^{n+1}-1\right) \left(q^{n+2}-1\right)^2 S_{1,2}(-\tfrac{1}{2}, 2; n; q)
\nonumber\\
&& +\left(q^{n+2}-1\right) \left(\left(q \left(q \left(\left(q^3+2 q-1\right)
   q^n-2\right)-3\right)
\right.\right.
\nonumber\\ && \left. \left.
+1\right)
q^{n+1}
+2\right)
   S_{1,2}(-\tfrac{1}{2}, 2; n+1; q)
\nonumber\\ &&
+\left(\left(((2-5 q)
   q+6) q^{n+2}+\left(q^2 (2
   q-1)-2\right) q^{2 n+4}
\right. \right.
\nonumber\\ && 
+4
   q 
\left. \left.
-7\right) q^{n+2}+1\right)
   S_{1,2}(-\tfrac{1}{2}, 2; n+2; q)
\nonumber\\ && 
-2 \left(q^{n+3}-1\right)^3
   S_{1,2}(-\tfrac{1}{2}, 2; n+3; q) = 0.
\end{eqnarray}
A $q$-differential equation for a $q$-extended harmonic polylogarithm is given by
\begin{eqnarray}
&& \biggl[{\cal D}_x^{(q)}
+q
   \left[q^2 x+q (3 x-1)+3
   x-2\right]
   ({\cal D}_x^{(q)})^2
\nonumber\\ &&
+\left[q^5 x^2+q^4 x (2
   x-1)+q^3 x (3 x-2) \right.
\nonumber\\ && \left.
-q^2 x
+q
   (2 x-1)-1\right]
   ({\cal D}_x^{(q)})^3
\nonumber\\ &&
+q^2 x (q
   x-1) \left(q^3 x+1\right)
   ({\cal D}_x^{(q)})^4 \biggr] {\rm H}_{-1,0,1}(x;q) 
= 0. \nonumber\\
\end{eqnarray}
Finally, an example for a $q$-shift equation reads
\begin{eqnarray}
&& q^4 (x-1) (x+1) F[x]   
  +q^2 \left(q^2 \left(-4 x^2+2 x+1\right) \right.
\nonumber\\ && \left.
-2 q (x-1)+1\right) F[q x]
\nonumber\\ &&
  +q \left(3 q^3 x (2 x-1)-2 q^2+q (3 x-2)-2\right) F\left[q^2 x\right]
\nonumber\\ &&
  +\left(q^4 \left(x-4 x^2\right)+3 q^3 x+q^2 (1-3 x) \right.
\nonumber\\ && \left.
-q
   (x-2)+1\right) F\left[q^3 x\right]
  +(q x-1) \left(q^3 x+1\right) 
\nonumber\\ && \times
F\left[q^4 x\right]
= 0,  
\end{eqnarray}
with $F[x] =  {\rm H}_{-1,0,1}(x;q)$.

An example for an iterated integral containing square-root valued letters is, 
cf.~\cite{Ablinger:2014bra},
\begin{eqnarray}
T_4(x) = \sum_{n=1}^\infty \frac{4^n}{\displaystyle n^2 \binom{2 n}{n}} S_1(n) x^n.
\end{eqnarray}
We consider the $q$-extension of $T_4(x)$.
{\footnotesize
\begin{eqnarray}
\label{T4xq}  
T_4(x;q) &=& \sum_{n=1}^\infty 4^n \left(\frac{1-q}{1-q^n}\right)^2
\frac{\displaystyle \prod_{k=1}^n
(1-q^k)^2}{\displaystyle \prod_{k=1}^{2n}
(1-q^k)}
\sum_{k=1}^n \frac{(1-q)}{(1-q^k)}
~x^n 
\nonumber\\
&\equiv& \sum_{n=1}^\infty f[n;q] x^n.
\end{eqnarray}
}

The associated $q$-differential equation reads
\begin{eqnarray}
&& \biggl\{
8 + Q_7 {\cal D}_x^{q}
+Q_8 ({\cal D}_x^{q})^2
+Q_9 ({\cal D}_x^{q})^3
+Q_{10} ({\cal D}_x^{q})^4
\nonumber \\ &&
+Q_{11} ({\cal D}_x^{q})^5
+Q_{12} ({\cal D}_x^{q})^6
+Q_{13} ({\cal D}_x^{q})^7   
\nonumber\\ &&
+(q-1)^4 q^{24} x^7 ({\cal D}_x^{q})^8 \biggr\} F[x;q] = 0. 
\end{eqnarray}
For the polynomials $Q_i$ see Ref.~\cite{Blumlein:2026shz}.
The corresponding $q$-shift relation is
\begin{eqnarray} &&
\label{eq:47a}
q^5 (1-4x)^2 F[x]
+ Q_1 F[q x]  
+ Q_2 F[q^2 x]
\nonumber\\ &&
+ Q_3 F[q^3 x]
+ Q_4 F[q^4 x]
+ Q_5 F[q^5 x]
\nonumber\\ &&
+ Q_6 F[q^6 x]  
+ (q-1) F[q^7 x]
+ F[q^8 x] 
\nonumber\\ &&
+ 8 (q-1)^4 q^4 x = 0. 
\end{eqnarray}
Structurally the finite binomial case is not different from the other function classes
in the case of the $q$-extension, unlike for the $\mu$-extension.
\section{Conclusions}
\label{sec:5}

\noindent
For the $\mu$-extension, the $\mu$-dependence emerges polynomially, leaving the
nested sums and iterated integrals otherwise in most of the nested sums and iterated 
integrals. Lower weight objects, down to $w = 0$, contribute.
This is different for {finite binomial sums} and iterated integrals 
over root-valued alphabets.
The $q$-extensions are different from the $\mu$-extensions.
The defining equations are $q$-difference equations for the nested $q$-sums and 
$q$-differential equations and $q$-shift relations for the $q$-extended iterative 
integrals.
In all function classes one has similar type defining equations.
For the simplest cases, one can calculate the $\mu$- and $q$-extensions for the 
whole class analytically. For more involved alphabets we have 
computer-algebraic algorithms to compute all individual cases, cf.~Refs.~\cite{SIG1,SIG2,
UNCU,KOUTSCHAN}. Also the $\mu$- and $q$-extended structures obey shuffle and stuffle
algebraic relations.
There are also solutions of {non-first order factorizing} 
differential and difference equations in the non-extended case in Quantum  Field Theory,
which have not been considered here.  

\vspace*{2mm}
\noindent
{\bf Acknowledgment.} This paper has been presented at the Bogolyubov Kyiv Conference 
`Problems of Theoretical and Mathematical Physics', Sept.~2026. Support from KMPB Berlin
is gratefully acknowledged for performing this work. A.M.G.was also funded by the 
National Academy of Sciences of Ukraine by its priority project No. 0122U000888.
C.S. has been funded in part by the
Austrian Science Fund (FWF) grants DOI 10.55776/P33530
and DOI 10.55776/PAT1332123.

\end{document}